# Advancement on Security Applications of Private Intersection Sum Protocol

Athur Raghuvir, Y., Govindarajan, S., Vijayakumar, S., Yadlapalli, P and Di Troia, F. [1]

[1] Computer Engineering San Jose State University
1 Washington Sq, San Jose, CA 95192, USA
`{yuva.athur, senthilanand.govindarajan, sanjeevi.vijayakumar, pradeep.yadlapalli, fabio.ditroia}@sjsu.edu`

**Abstract.** Secure computation protocols combine inputs from involved parties to generate an output while keeping their inputs private. Private Set Intersection (PSI) is a secure computation protocol that allows two parties, who each hold a set of items, to learn the intersection of their sets without revealing anything else about the items. Private Intersection Sum (PIS) extends PSI when the two parties want to learn the cardinality of the intersection, as well as the sum of the associated integer values for each identifier in the intersection, but nothing more. Finally, Private Join and Compute (PJC) is a scalable extension of PIS protocol to help organizations work together with confidential data sets. The extensions proposed in this paper include: (a) extending PJC protocol to additional data columns and applying columnar aggregation based on supported homomorphic operations, (b) exploring Ring Learning with Errors (RLWE) homomorphic encryption schemes to apply arithmetic operations such as sum and sum of squares, (c) ensuring stronger security using mutual authentication of communicating parties using certificates, and (d) developing a Website to operationalize such a service offering. We applied our results to develop a Proof-of-Concept solution called JingBing, a voter list validation service that allows different states to register, acquire secure communication modules, install it, and then conduct authenticated peer-to-peer communication. We conclude our paper with directions for future research to make such a solution scalable for practical real-life scenarios.

## 1  Introduction

Data Collaboration is a growing challenge in a world that is getting increasingly digital. Organizations are now having loads of information from which value can be generated supporting more data driven decisions. In many scenarios, collaborating organizations have common identifiers and data associated with common identifiers from which knowledge can be gleaned. These identifiers can range from innocuous identifiers of products to personally identifiable information of people. To these identifiers, meaningful data can be associated.

In secure multi-party computation, each party possesses some private data, while secret sharing provides a way for one party to spread information on a secret such that all parties together hold full information, yet no single party has all the information. There are several interesting real-life scenarios that can benefit from secure multi-party communication [1][3]. For example, Credit Card Number that uniquely identifies a person can be associated with their spending patterns in a retail organization and can be associated with use of transit in the state transportation department. By offering a way to combine a retailer's data set that consists of spending patterns to the same user's transit usage, information about how many visitors are spending time in that retail organization can be discovered [24]. Similarly, manufacturers associate product Stock Keeping Unit (SKU) with product specification and production plans, while corresponding SKUs can be associated with consumption patterns at a distributor site. By combining these pieces of information, a manufacturer can perform real-time demand forecasting with real data as opposed to derived data from past years [1][2][10][12][13][14][15][16][17][24].

There were four main objectives for this research work:
- Understand in detail how Google's open-source Private Join and Compute (PJC) protocol works.
- Identify extensions to the PJC protocol to harden the communication between the communicating parties.
- Explore integration options with fully homomorphic encryption algorithms to extend the scope of operations supported by PJC in one protocol run.
- Apply the use of PJC to a real-life scenario and investigate mechanisms to take such a solution to market.

In this paper, we report our results on the above-mentioned objectives. In particular,
- We extended PJC protocol to transmit more data during a single run between client and server. At the same time, we were able to keep the server process listening across multiple protocol runs.
- Using certificates, we were able to further harden the identity and authentication of participants using the protocol by extending the gRPC implementation on which PJC is based.
- We were able to utilize the Microsoft SEAL Multiplicative homomorphic encryption library with the PJC protocol to generate sums of squares of encrypted data.
- We developed and deployed a prototypical website called JingBing, a voter list validation service that allows different states to register, acquire secure communication modules, install it, and then conduct authenticated peer-to-peer communication.

In the following sections we elaborate on the requirements for data collaboration at Scale, application of PJC to Voter List Validation problem, explain architecture, design and a Proof-of-Concept implementation of JingBing Service demonstrating the ideas in this paper, and finally close with our observations and directions for future research.

## 2 Requirements for Data Collaboration at Scale

We investigated the challenges and requirements to support data collaboration at scale [12][13]. During our investigation, we have identified the following key aspects that need to be addressed to solve Data Collaboration across organizations.

**Public Communication Channel** - Most secure communications come with a need for a preestablished secure channel limiting the possibility of data collaboration with only pre-established partners. Ad-hoc collaboration is typically not possible.

**Intellectual Property protection -** Organizations would like to know information that the other organizations have without having to divulge the information they have to protect the data IP that the organizations have.

**Requirement of Specific metrics and measures -** Since sharing data is not preferred due to data IP issues, the next best level of information that can be shared are derived metrics from the data. For many real-life scenarios, generating appropriate metrics and measures from other party's data is good enough to lead to data-driven decisions.

**Performance and scalability -** As data collaboration across organizations becomes a key aspect of the digital supply chain, it is important to consider performance and scalability of these operations. So, any solution proposed must offer desired performance and scalability for that use case.

Secure computation protocols allow parties to combine their inputs to generate output while keeping privacy of the inputs. [1][3][14] The specific family of data collaboration problems we are investigating gives rise to Intersection-Sum Problem. In this abstraction, there are two activities with two different parties. The first party has a list of identifiers associated with the first activity, and the second side has a list of identifiers that transacted on a second activity. One party wants to learn the number of identifiers they have in common, and the sum of some associated data that the identifier generates. By having a secure data collaboration protocol, we learn how many identifiers of first activity drive towards the second activity, and secondly, we learn how much the first activity increases the key indicators (like sum or average) in the second activity. Solutions for Intersection-Sum problems using secure computations start from Private Set Intersection protocols. [12][13][14] Private Set Intersection (PSI) is a secure computation protocol that allows two parties, who each hold a set of items, to learn the intersection of their sets without revealing anything else about the items. Private Intersection Sum (PIS) extends PSI when the two parties want to learn cardinality of the intersection, as well as the sum of the associated integer values for each identifier in the intersection, but nothing more [13][14].

Finally, Private Join and Compute (PJC) is a scalable extension of PIS protocol to help organizations work together with confidential data sets. [24]

## 3   Problem and Motivation: Extending PJC to Voter List Validation Problem

Google has studied various ways of making PJC performant and scalable [14]. Furthermore, Google has open sourced a simple client-server protocol that demonstrates how such a PJC protocol could be implemented [8]. In this paper we have taken a deeper look into Google's PJC, improving its implementation, and applied it to Voter List Validation.

### 3.1   Voter List Validation Problem

States allocate budgets based on population distribution. [11][22] Voters database is one such source of data. Voters registered in multiple states leads to over-allocation of resources and undue financial burden on the states. Also, due to the increased concerns around voter fraud, states are required to find out whether there are any voters registered in two or more states.

Collecting the data indicating the number of common voters and aggregating the financial dollars allocated to these voters gives visibility to states to analyze the impact of individuals registered in multiple states. However, it is important to hide individual identity due to Federal and State privacy regulations. Thus, there is a need for a solution that is apt but also preserves individual privacy thereby revealing only the minimum requested criteria that includes information about the number of common voters along with the aggregated value impacted by them.

### 3.2   JingBing

We have developed a Voter List Validation Service called JingBing. JingBing uses Private Join and Compute (PJC) which is a secure communication protocol using homomorphic encryption that allows the parties in communication to compute the other party's encrypted data without decrypting it. Here, the data between the parties is exchanged multiple times to encrypt the data with both party random keys ensuring data security.

It is important to shuffle the data in the exercise to maintain the data position random, to avoid revealing the identity of voters that are common in both datasets. [14] In order to apply PJC to JingBing, the protocol must be extended to restrict communication only between known parties who are mutually authenticated through JingBing certificates. Further, invalid datasets pattern attempts from Semi-honest or Honest- but-curious participants will be identified by the protocol. Protocols that provide security against "honest-but-curious" adversaries, assume that the participants will follow the protocol steps honestly, but may try to learn as much as possible from the protocol messages. Honest-but-curious security means that such participants should learn nothing more than the protocol prescribes. On the other hand, malicious adversaries can deviate arbitrarily from the protocol, and malicious security for a construction guarantees that even such adversaries cannot learn more about the private inputs.[14]

The need for identifying multi-state registration has been long identified as an issue that needs to be addressed. Interstate voting is considered a felony in many states and when identified, the individual is imposed hefty fines and jail time. Currently, there is no targeted or focused solution to compute the resources or financial dollars spent on the Voters by each state.

## 4   Prior Work: Crosscheck Program

In December 2005, the office of Kansas Secretary of State initiated a database software program named Interstate Voter Registration Crosscheck or Crosscheck in short [4]. This program was targeted to identify duplicate registrations by voters in multiple states by comparing the first name, last name, date of birth, and the last four digits of the SSN. Implementation of Crosscheck revealed flaws and limitations as follows [4][5][23]:

**Inaccurate Results** - This technique yields inaccurate results where the last four digits of the SSN are not either matched or went missing. These inaccurate results led to the inappropriate deletion of eligible voters in many counties.

**Data security and Data handling issues -** There are concerns raised about the opportunity for racial bias with the information sharing between the states where there are lapses in data security and data handling. Most of the non-public information like SSN number, date of birth, etc. were shared across the states.

**Limits to Crosscheck functionality** - This program cannot detect the double voting within a state. Also, waiting on behalf of a deceased person cannot be detected.

**Vulnerable Security protocols -** In fall 2017, it was discovered that the Kansas-managed database holds nearly 100 million records of private voter data. However, the security protocols were identified to be extremely vulnerable making it an easy target for hackers. [19] For example, Crosscheck's files are hosted on an insecure server, according to its own information. Usernames and passwords were regularly shared by email, making them vulnerable to snooping. Furthermore, the passwords were overly simplistic and only irregularly changed.

**Poor management and transfer of records -** Further investigation into mechanisms of data transfer revealed that username and password sent to the program's neglected to redact before releasing the emails publicly. Unencrypted file transfer protocols were used. In cases where encrypted data was used the passwords to decrypt three years' worth of voter files, belonging to every state participating in Crosscheck, have likewise been exposed. In December 2019, as part of a settlement of a lawsuit filed by ACLU of Kansas challenging Kansas' management of the program, the Crosscheck program was suspended indefinitely.[5]

Based on the experience of the Kansas Crosscheck program, we were able to establish that the Voter Validation Service proposed is a real-life scenario that could be solved using an extension of the Private Join and Compute (PJC) protocol.

Our contribution has been to extend PJC as a secure communication protocol to offer the following benefits:

**Peer-to-Peer Communication -** PJC fundamentally eliminates the need for a middleman. This is a significant step forward in securing any data collaboration.

**Privacy-preserving on Public internet -** The use of cryptographic methods offers a way to set up peer-to-peer data exchanges without having to set up secure communication channels.

**Multiple measures and metrics -** Using the extension of the PJC protocol proposed in this project, the scope of meaningful data that can be shared can be extended. This would be based on the scope of homomorphic encryption functions that would be supported in the extension of PJC.

**Authenticated communication -** Adding an authentication envelope to the PJC protocol enables computation between trusted parties.

## 5   JingBing Architecture and Design

### 5.1   Introduction

JingBing Voter list validation service is a Proof of Concept (PoC) to demonstrate the viability of our solution. JingBing enables two states holding their voter lists to identify the double-registered voters among them and compute the accrued expenses without revealing the identities of those voters. A website will provide the states with services required to utilize the proposed system.

The software bundle delivered by the website includes a secure computation engine and an authentication module. The downloaded software bundle allows states to generate statistical inferences on different expenses data while providing data privacy and confidentiality.

## 5.2 Architecture of Subsystems

The schematic diagram of all the subsystems that participate in JingBing solutions is illustrated in Fig. 1.

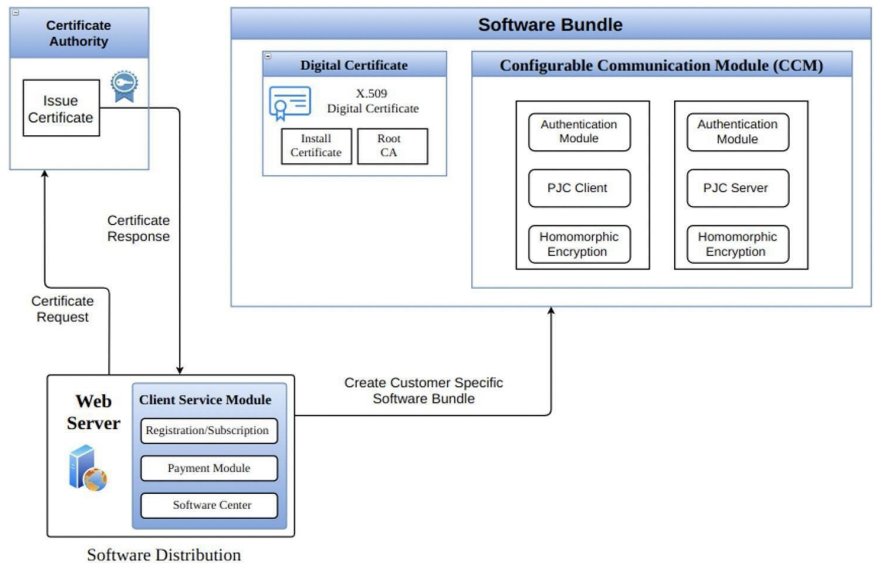

**Fig. 1.** System Architecture

The main components of the New Generation Voter list validation service are:
- Website
- Mutual Authentication Module
- Configurable Communication Module (CCM)
- Homomorphic Encryption Scheme Selector

JingBing website is designed for US States to register and participate in the data exchange. Users representing a state are required to fill mandatory fields that are used for validating the user and the entity they represent. Once registered, the user can browse through a directory of states that have active subscriptions with the service.

Each user must generate and import a certificate from JingBing. At the start of the session, the parties must share their certificates to establish authenticity. Without this mutual authentication, the protocol will not proceed. Once the two participating parties are authenticated, the next phase is to compute the desired computation on the intersecting data points of the two data sets.

The data exchange protocol is managed through a communication module that allows the parties to participate as a client or a server. Client initiates the protocol. Through a sequence of controlled interactions as defined in the PJC protocol structure, the data is exchanged over public internet. The multi-step data exchange ensures secure privacy preserving collaboration in the following ways:
- The intersecting set of records are determined using commutative cipher encryption technique.
- There is shuffle step explicitly included to avoid any malicious actor infer any information due to preserved order.
- Client requests the Server to perform a certain computation on the intersecting set of data elements.
- Homomorphic encryption ensures that requested arithmetic operations can be performed on encrypted data.
- The summarized results are then passed back to the client.

The next section describes the implementation details.

# 6 Proof-of-Concept Implementation

The JingBing web server serves as the public website through which organizations desiring to have a secure data exchange using this novel protocol can register, acquire, configure, and run the protocol. Smart mutual authentication eliminates the need for middleman during the data exchange. This is critical for maintaining the privacy of the data exchange. Furthermore, the use of homomorphic encryption schemes guarantees that honest and semi-honest participants of the protocol will not be able to learn anything more than what is intended by the protocol. Finally, the extension of the PJC protocol using CCM, along with addition of homomorphic encryption schemes and mutual authentication mechanism, had been developed to demonstrate how this protocol would work in practice.

## 6.1 Delivery and Subscription Website

The delivery and subscription of the product is done through its website. The Website was built on a website-as-a-service platform, Wild Apricot [26]. The implementation of the JingBing website can be schematically shown as in Fig. 1, and it is split in three main sections:
1. **Web Server.** The main website that provides functionality of registration of users, downloads of software components, requesting certificates for mutual authentication, listing of other members, payment options and support.
2. **Software Bundle.** The software package that consists of a certificate and a Configurable Communication Module for secure data exchange between the parties.
3. **Certificate Authority.** When the party requests for a certificate, it is issued from a Certificate Issuing Authority.

### 6.2 Mutual Authentication Module

The mutual authentication [25] module for the PJC involves.

(a) **Extending gRPC framework** [9] for mutual key-based authentication and
(b) **Public Key Infrastructure (PKI)** authority to issue certificates for the participant states

PJC protocol utilizes gRPC as the transport layer for its operations. The gRPC framework was extended to include a key-based mutual authentication scheme. For a successful connection, the client needs to present a certificate that can be verified against the root certificate configured by the server.

Vault PKI [20] is an open-source secrets management software which enables very strong authentication and a policy-based access control foundation for managing digital secrets.

A policy-enabled standalone instance of Vault PKI secrets engine was configured for providing secure certificates, and these certificates were used to mutually authenticate participants of the protocol.

### 6.3 Configurable Communication Module (CCM) Implementation

We extended open-source PJC protocol to support the following two goals:

a) **Adding another data column** to the protocol
b) **Including one more operation** other than the sum that can be requested from the client to the server

To achieve these goals, three main extensions were introduced:

1. **Protocol Extension.** PJC protocol was extended to support sharing multiple data columns.
2. **Client Module Extension.** The Client Module was extended to read multiple columns, take flags for multiple operations, and exchange data with the server appropriately.
3. **Server Module Extension.** The Server Module was extended to receive operator instructions from the client, perform the required operations using homomorphic library operators, and return the computed result to the client.

In these extensions the PJC protocol structure does not change. The operators are passed to the client call along with the data. Schematically this can be represented as shown in Fig. 2.

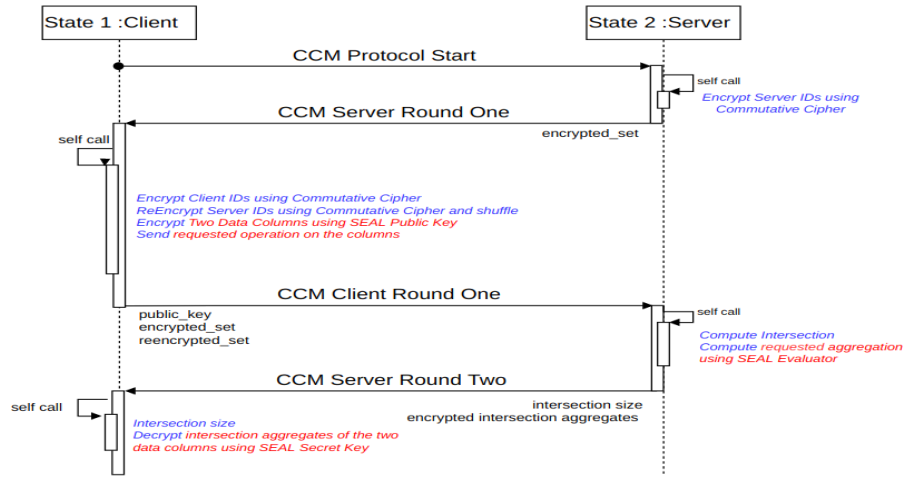

**Fig. 2.** Sequence diagram between Client and Server

### 6.4 Extensions to Homomorphic Encryption Schemes

PJC protocol utilizes Paillier partial homomorphic encryption library for sum operations [8]. Paillier partial homomorphic encryption library supports only homomorphic sum operations and does not support homomorphic multiplication operations [14]. Fully homomorphic encryption capability was added to PJC through integration of Microsoft SEAL [18]. Microsoft SEAL is a fully homomorphic encryption library providing both homomorphic additive and multiplicative capabilities. The BFV Scheme [6][18] of Microsoft SEAL was used in JingBing application for integration since it is applicable for real numbers and results in accurate decryption of ciphertext. Microsoft SEAL was statically linked to CCM code directly. Pluggable On-demand Modular Homomorphic encryption engine using Microsoft SEAL library was also developed [21].

## 7 Summary, Conclusions, and Recommendations

### 7.1 Summary

In this paper, we have taken a deeper look into Google's Private Join and Compute protocol and applied it to the Voter List Validation problem. The key contributions are:

- **JingBing Webservice for Voter List Validation** as a proof of concept on viability of our solution.
- **Mutual Authentication mechanism** using Mutual Transport Layer Security (mTLS).

- **Implemented Configurable Communication Module** extending Google's PJC protocol to support sharing multiple data columns and additional aggregation operations on data columns.
- **Utilized Microsoft SEAL Fully homomorphic encryption** to accommodate more aggregation operations on data columns.

**JingBing WebService for Voter List Validation**
To take this idea to produce quality service for a real-life problem, the JingBing Voter List Validation service was implemented to closely prototype client registration, client authentication, certificate procurement, software procurement, and tiered pricing model. The PoC implementation has the following features using WildApricot Website as a service platform:
- A subscription was enabled with mandatory input fields that tie every account to one of the U.S. states.
- Directory of active states with membership is shared through the website so they can interact and participate in extended PJC secure data exchange protocol.
- Generation and issue of authentication certificate using Vault PKI integration.
- Software download has been enabled with the latest bundle file uploaded to the account and security level added to download member-only.
- Emails can be sent by users to the support team for any issues with website usage. Details are available on the Contact Us page.

**Mutual Authentication mechanism using mTLS**
One key observation we made when we embarked on extending the PJC protocol for real-life scenarios is the lack of a strong mutual authentication approach. To eliminate the need for an intermediate 3rd party during the protocol run, we came up with a Mutual Transport Layer Security(mTLS) based solution. The key features that we implemented as part of mTLS are:
- Vault PKI [20] was used as a Web-enabled PKI Authority to issue digital certificates to be used for authenticating participant states.
- PKI was hardened with a token-based authentication and policy-based authorization scheme.
- gRPC Framework development for mTLS Extension to authenticate participant states.
- Logging was enabled to demonstrate non-repudiation during data exchange in one protocol run.

**Extended PJC protocol to support sharing multiple data columns.**
PJC protocol is a promising new protocol that enables privacy-preserving computation over the open internet using smart encryption techniques like commutative ciphers and homomorphic encryption. To make this protocol scalable to real-life scenarios, we implemented Configurable Communication Module [7] that accomplished the following:
- Supports multiple columns by extending the message content in the protocol.

- Computed intersection sum and intersection sum of squares of data elements.
- Extend the sample data set generation to include generating multiple columns and apply different operations.
- New flags to drive protocol behavior from the command-line. This gives more flexibility to the client and server communication.
- Embedded mTLS into the protocol to keep the entire protocol safe.

**Integrated Microsoft SEAL homomorphic encryption to accommodate more operations on data columns.**

The available open-source implementations of Fully Homomorphic Encryption technologies like Microsoft SEAL, Google SHELL, and ElGamal were studied for integration into PJC protocol implementation. For the MVP implementation, the following were accomplished:
- Successfully integrated Microsoft SEAL Homomorphic encryption library
- Used Microsoft SEAL's BFV scheme to perform sum and sum of squares operations to extend the homomorphic operations.

### 7.2  Conclusions

Based on our research, there is currently no viable solution that can handle the Voter List Validation problem efficiently while preserving the security and privacy of participants' data. The last known solution called CrossCheck implemented in Kansas state has been indefinitely suspended since 2019 due to lack of data security issues and inaccurate results. This led us to choose Voter List Validation as a problem to apply the extension of PJC protocol for data exchange between states.

Our JingBing service is a novel solution and solves the problems that States face when they must share data in a secure and privacy preserving manner. With JingBing they can fulfil the need for a decentralized solution over open internet communications. States that participate can now initiate a trusted and secure data exchange protocol with the other States whenever they see the need for determining common sets of voters and find associated business KPIs.

The current Proof-of-Concept implementation of JingBing Voter List Validation Service has accomplished the following key points:
- Identification of intersection size without revealing the personal identity of the persons involved.
- Share additional data columns to offer more insights into how states can exchange computations on data.
- Hardened PJC protocol to include key-based mutual authentication.
- Use aggregation operators like sum and product to get deeper insights on the data to drive data-driven decision making for the states participating in the protocol.

With JingBing Voter Validation Service PoC, we have successfully explored the application of secure privacy-preserving intersection sum protocol. We have further established that the crypto techniques have come of age for use in real-life applications.

### 7.3 Recommendations for Further Research

JingBing Voter Validation Service encapsulates how a secure protocol can be productized for a real-life scenario. Below are some topics for future research:

- Extend Mutual Authentication to Token-based certificate handling.
- Introduce responsible logging mechanisms satisfying Principles of private computing to support audit and non- repudiation requirements.
- Thinking of tags as database table columns offer some interesting options of columnar analytics expressed in SQL syntax with possible extensions that are borrowed from modern DWH systems like BigQuery.
- The current implementation of CCM with SEAL integration has low limits on value bounds of the column (31) and number of records that can be transmitted in one run (20). The first constraint is coming from the parameters used to initialize SEAL BFV FHE scheme and the second constraint is coming from gRPC internal limits, respectively. Extending to larger values and more records are required to address practical scenarios where data set sizes and value of data is expected to be larger.
- The hardness of the underlying cryptographic procedures has a performance downside. More work needs to be done to improve performance without compromising on the security and privacy preserving nature of the protocol.
- Google's Private Join and Compute (PJC) protocol is a simple, robust, and scalable protocol for secure data exchange using advanced cryptographic techniques. In this paper, we applied our solution to a Proof of Concept (PoC) that demonstrates how such a protocol can be taken forward to implement a real-life Voter List Validation Service ensuring privacy-preserving secure data exchange. The current PoC offers a software bundle that validates passing 2 data columns and allowing different operators. Based on the above suggestions, the Voter List Validation Service can be improved to develop a more robust solution.

## References

1. Archer, W.D., Bogdanov. D., Kamm. L., Lindell. Y., Nielsen. K., Pagter. I.J., Smart. P.N., Wright N.R. (2018). From Keys to Databases -- Real-World Applications of Secure Multi-Party Computation, https://eprint.iacr.org/2018/450.pdf, last accessed 2021/03/15.
2. Buddhavarapu, P., Knox, A., Payman Mohassel, Shubho Sengupta, Erik Taubeneck, Vlad Vlaskin, Private matching for compute: New solutions to the problem of enabling compute on private set intersections, https://engineering.fb.com/2020/07/10/open-source/private-matching/ , last accessed 2021/03/15.
3. Cramer, R., Damgård, I., & Nielsen, J. (2015). Secure Multiparty Computation and Secret Sharing. Cambridge: Cambridge University Press. doi:10.1017/CBO9781107337756\

4. Cross-check program, https://www.endcrosscheck.com, last accessed 2021/03/15.
5. Cross check program suspension, https://www.aclukansas.org/en/press-releases/aclu-kansas-settlement-puts-crosscheck-out-commission-foreseeable-future-program, last accessed 2021/03/15.
6. Fan, J., and Vercauteren, F., (2012) Somewhat Practical Fully Homomorphic Encryption, https://eprint.iacr.org/2012/144, last accessed 2021/03/15.
7. Github.com (2020) Configurable Communication Module extending Google's PJC, https://github.com/YuvaAthur/private-join-and-compute, last accessed 2021/03/15.
8. Github.com (2019). Private join and compute, https://github.com/google/private-join-and-compute, last accessed 2021/03/15.
9. gRPC, https://grpc.io/, last accessed 2021/03/15.
10. Giordano, G., Powering larger insights with privacy-preserving computation, https://www.accenture.com/us-en/blogs/technology-innovation/giordano-linvill-privacy-preserving-computation, last accessed 2021/03/15.
11. How Will States and Localities Divide the Fiscal Relief in the Coronavirus Relief Fund? https://www.cbpp.org/research/state-budget-and-tax/how-will-states-and-localities-divide-the-fiscal-relief-in-the, last accessed 2021/03/15.
12. Ion, M., Kreuter, B., Erhan Nergiz, A., Patel. S., Raykova, M., Saxena, S., Seth, K., Shanahan, D. and Yung, M. (2017). Private Intersection-Sum Protocol with Applications to Attributing Aggregate Ad Conversions, https://eprint.iacr.org/2017/738.pdf, last accessed 2021/03/15.
13. Ion, M., Kreuter, B., Erhan Nergiz, A., Patel. S., Raykova, M., Saxena, S., Seth, K., Shanahan, D. and Yung, M. (2019). On Deploying Secure Computing: Private Intersection-Sum-with-Cardinality, https://eprint.iacr.org/2019/723.pdf, last accessed 2021/03/15.
14. Ion, M., Kreuter, B., Erhan Nergiz, A., Patel. S., Raykova, M., Saxena, S., Seth, K., Shanahan, D. and Yung, M. (2019). On Deploying Secure Computing Commercially: Private Intersection-Sum Protocols and their Business Applications, https://eprint.iacr.org/2019/723/20190618:153102, last accessed 2021/03/15.
15. Lawton, G., Confidential computing promises secure cloud apps, https://searchcloudcomputing.techtarget.com/tip/Confidential-computing-promises-secure-cloud-apps, last accessed 2021/03/15.
16. Lepoint, T., Patel. S., Raykova, M., Seth, K., Trieu. N. (2021). Private Join and Compute from PIR with Default, https://eprint.iacr.org/2020/1011.pdf, last accessed 2021/03/15.
17. Miao, P., Patel. S., Raykova, M., Seth, K. and Yung, M. (2020). Two-Sided Malicious Security for Private Intersection-Sum with Cardinality, https://eprint.iacr.org/2020/385.pdf, last accessed 2021/03/15.
18. Microsoft SEAL library, https://www.microsoft.com/en-us/research/project/microsoft-seal/, last accessed 2021/03/15.
19. More vulnerabilities in crosscheck, Avail- able: https://gizmodo.com/even-a-novice-hacker-could-breach-the- network-hosting-k-1820263699, last accessed 2021/03/15.
20. PKI Secrets Engine, https://www.vaultproject.io/docs/secrets/pki, last accessed 2021/03/15.
21. Sealdev-github, https://github.com/vsanjeevi/SealDev, last accessed 2021/03/15.
22. These Are Our Numbers – The Importance of the 2020 Census, https://budget.house.gov/publications/report/these-are-our-numbers-importance-2020-census, last accessed 2021/03/15.
23. Vulnerabilities in crosscheck, https://www.propublica.org/article/crosscheck-thevoter-fraud- commission-wants-your-data-keep-it-safe, last accessed 2021/03/15.

24. Walker, A., Patel, S., and Yung, M., Blog post Helping organizations do more without collecting more data, https://security.googleblog.com/2019/06/helping-organizations-do-more-without-collecting-more-data.html, last accessed 2021/03/15.
25. What is Mutual Authentication, https://community.developer.visa.com/t5/Developer-Tools/What-is- Mutual-Authentication/ba-p/5757, last accessed 2021/03/15.
26. Wild Apricot: Website as a Service, https://www.wildapricot.com/ , last accessed 2021/03/15.